\documentclass[11pt]{article}

\usepackage{amsmath,amssymb,amsthm,graphicx,latexsym}
\newcommand{\ed}{\end{document}}
\newcommand{\beq}{\begin{equation}}
\newcommand{\eeq}{\end{equation}}
\newcommand{\beqa}{\begin{eqnarray}}
\newcommand{\eeqa}{\end{eqnarray}}
\newcommand{\bc}{\begin{center}}
\newcommand{\ec}{\end{center}}

\newcommand{\ba}{\begin{array}}
\newcommand{\ea}{\end{array}}
\newcommand{\pa}{\partial}

\begin{document}
%%%%%%%%%%%%%%%%%%%%%%%%%%%%%%%%%%%%%%%%%%%
%%%%%%%%%%%%%%%%%%%%%%%%%%%%%%%%%%%%%
\title{{\bf{Canonical formulation for nonrelativistic nonisentropic Euler fluids and Schwinger-type conditions }}}
%%%%%%%%%%%%%%%%%%%%
\author{  {\bf {\normalsize Rabin Banerjee$^1$}$
$\thanks{E-mail: rabin@bose.res.in}},$~$
{\bf {\normalsize Arpan Krishna Mitra$^1$}$
$\thanks{E-mail: arpan@bose.res.in}}$~$
\\{\normalsize $^1$S. N. Bose National Centre for Basic Sciences,}
\\{\normalsize JD Block, Sector III, Salt Lake, Kolkata-700098, India}
\\\\
\\[0.3cm]
}
\date{}
\maketitle

\begin{abstract}

We present a new approach, based on Noether's energy-momentum tensor, to construct the lagrangian for nonrelativistic nonisentropic Euler fluids. An advantage of this approach is that it naturally provides a generalised Clebsh decomposition for the fluid velocity. This is used to develop a hamiltonian formulation inolving a noncanonical algebra. This algebra is very simply obtained from the symplectic structure. It is used to show that the components of the Noether's energy-momentum tensor satisfy certain Schwinger-type relations. These relations, which are reminiscent of corresponding relations in relativistic field theory, are new.
\end{abstract}
\pagebreak
\section {Introduction}
Despite the fact that the origins of fluid dynamics lie in nineteenth century physics, its relevance has not abated even now. It has stood the test of time like Maxwell's electrodynamics. The theory of fluid dynamics has continuously evolved through its various ramifications and extensions. Indeed it can and does illuminate several features of particle physics, especially those related to extended structures, and also gravity, that has culminated in the fluid/ gravity correspondence\cite{hub}. Thus the study of fluid dynamics, which is interesting in its own right, has relevance and significance in the modern context.

The dynamics of fluids is described by a classical field theory using either the lagrangian or hamiltonian approach. The corresponding variables are the Euler variables since fluid dynamics is most conveniently expressed in terms of them \cite{lan, hol, gre}. Among these competing approaches, that based on the hamiltonian is quite frequently discussed. This is not surprising as the form of the hamiltonian may be written on general principles and the algebra of variables suitably defined to reproduce the known equations of motion for the fluid \cite{gre, jac}. Incidentally, this algebra is either posited by inspection \cite{gre} or derived by using the Lagrange to Euler map \cite{avi}. A self contained derivation within the Eulerian scheme seems to be lacking.

The lagrangian formulation, on the contrary, is quite tricky. For usual canonical brackets the passage from the hamiltonian to the lagrangian is smooth, using an appropriate Legendre transform. However  when the brackets are non canonical, as happens in the case of fluids, this transition is far from straightforward. Indeed, it has been explicitly shown that the presence of Casimirs is the root of the obstruction \cite{baz, jac}. Under these circumstances, something  different has to be done. A possible way is to introduce Clebsh variables\cite{cl} and use certain continuity equations. This was the original method adopted by Lin and Eckart \cite{kara}. However, there were ambiguities since the method was ad-hoc and there was no proper physical basis for the choice of one continuity equation \cite{jac}. Another approach based on conservation laws is discussed in \cite{ff} but it also suffers from similar criticisms.

 In the present paper we provide a new Lagrangian approach to discuss non-relativistic fluids. We exploit Noether's definition\footnote{We recall the use of Noether's definition for relativistic hydrodynamics \cite{koda}. However, since Noethers definition is asymmetric, it has to be used with care when treating relativistic systems where the energy-momentum tensor must be symmetric and suitable improvements have to be done. In the non-relativistic case, of course,  this  restriction of symmetricity(between space-time) no longer holds.} of the stress tensor to obtain the cherished lagrangian. Ambiguities are avoided in this derivation. The choice of the Clebsh parametrisation is naturally dictated by the analysis. Nonisentropic fluids have also been considered. A generalised Clebsh parametrisation involving entropy is found which is a new result. A hamiltonian formulation has been given where non canonical brackets are computed directly from the symplectic structure. Exploiting these brackets we then show that the components of the stress tensor satisfy an algebra that is highly reminiscent of Schwinger \cite{nger} conditions found in relativistic field theory. As far as we understand these relations have never been discussed in the literature on non-relativistic fluid dynamics. It is easy to prove the conservation of the stress tensor from these relations.

\section{Basic formalism} 
 Let us first briefly recapitulate the basic tenets of non-relativistic isentropic Eulerian fluids. The hamiltonian is given by, 

\begin{eqnarray}
\label{h}
H=\int dx (\frac{1}{2}\rho v^{2}+V(\rho))
\end{eqnarray}
where $\rho$ and $v_{i}$ are the fluid density and velocity, respectively. The fundamental fluid equations, namely the Euler equation and continuity equation, are reproduced by appropriate bracketing with (\ref{h}) by exploiting the following algebra \cite{gre, jac},
\begin{equation}
\label{two}
\{\rho(x), \rho(x')\}=0, ~~\{\rho(x), v_{i}(x')\}=\pa_{i}\delta(x-x'),~~~\{v_{i}(x),v_{j}(x')\}=-\frac{\omega_{ij}}{\rho}\delta(x-x')
\end{equation}
where,
\begin{equation}
\label{vor}
\omega_{ij}=\pa_{i}v_{j}-\pa_{j}v_{i}
\end{equation}

is the vorticity of the fluid. The continuity equation is reproduced as,
\begin{equation}
\label{pppp}
\dot{\rho}=\{\rho, H\}=\pa_{i}(\rho v_{i})
\end{equation}
Likewise the Euler equation is obtained as\footnote{suffix on $V$ implies a derivative; $V_{\rho}=\frac{\pa V}{\pa \rho}$},
\begin{equation}
\label{rrrr}
\dot{v_{i}}=\{v_{i}, H\}=v_{j}\pa_{j}v_{i}+\pa_{i}V_{\rho}(\rho)
\end{equation}
Note that the second term on the right side of (\ref{rrrr}) may be expressed in a familiar form by recalling the definition of pressure $P$ as a Legendre transform of V \cite{jac},
\begin{equation}
\label{axyz}
P(\rho)=\rho V_{\rho}-V(\rho)
\end{equation}
so that,
\begin{equation}
\label{press}
\frac{1}{\rho}\pa_{i}P=\pa_{i}V_{\rho}
\end{equation}
It is thus the pressure gradient which is consistent with the fact that we are discussing ideal hydrodynamics.

 Introducing the current $j_{i}=\rho v_{i}$ it is simple to find,
\begin{equation}
\nonumber
\{j_{i}(x),\rho(x')\}=\rho(x) \pa_{i}\delta(x-x')
\end{equation}
\begin{equation}
\label{qqq}
\{j_{i}(x),j_{k}(x')\}=j_{k}(x) \pa_{i}\delta(x-x')+j_{i}(x') \pa_{k}\delta(x-x')
\end{equation}
In order to construct an appropriate Lagrangian let us introduce a variable $\theta$ which is canonically conjugate to $\rho$. Then we may write the lagrangian density as,
\begin{equation}
\label{eee}
{\cal{L}}=\rho\dot{\theta}-{\cal{H}}=\rho\dot{\theta}-(\frac{1}{2}\rho v^{2}(\theta)+V(\rho))
\end{equation}

Here $\theta$ is not an independent variable. It is related to the fluid velocity $v_{i}$. The velocity will be expressed as a function of $\theta$ so that in the above Lagrangian $\rho$ and $\theta$ are the only variables of variation. It is of course necessary to abstract the connection of the velocity with the variable $\theta$ before it is possible to show that the correct hydrodynamical equations follow from (\ref{eee}). This is shown below.

To understand the meaning of $\theta$ in terms of the fluid variables we take recourse to Noether's definition of stress tensor
\begin{equation}
\label{uuu}
T^{\mu\nu}=\frac{\partial {\cal{L}}}{\pa(\pa_{\mu}F)} \pa^{\nu}F-{\cal{L}}g^{\mu\nu}
\end{equation}
where $F$ generically denotes the variables in the lagrangian.

The momentum density in a non relativistic theory coincides with the current,
\begin{equation}
\label{cur}
T^{0i}=j^{i}=\rho v^{i}
\end{equation}
Computing $T^{0i}$ from (\ref{eee}) and (\ref{uuu}), we get,
\begin{equation}
\label{iii}
T^{0i}=\rho\pa^{i}\theta
\end{equation}
which, equated to (\ref{cur}), yields the identification,
\begin{equation}
\label{iit}
v^{i}=\pa^{i}\theta
\end{equation}
Since the vorticity (\ref{vor}) vanishes, this corresponds to an irrotational fluid.

The correspondence (\ref{iit}) is algebraically consistent with (\ref{two}) because,
\begin{equation}
\label{tw}
\{\rho(x), v_{i}(x')\}=\{\rho(x), \pa_{i}\theta(x')\}=\pa_{i}\delta(x-x')
\end{equation}
recalling that $(\rho,\theta)$ are a canonical pair, as seen from the first order lagrangian density (\ref{eee}),
\begin{equation}
\label{hii}
\{\theta(x), \rho(x') \}=\delta(x-x')
\end{equation}
 The $v_{i}-v_{j}$ bracket following from (\ref{iit}) vanishes thereby reproducing the algebra (\ref{two}) for an irrotational fluid.

To complete the demonstration of the validity of (\ref{iit}) we show that it reproduces the energy flux $T^{i0}$ and stress tensor $T^{ij}$ from (\ref{uuu}).

The first step is to express the lagrangian (\ref{eee}) in terms of $(\rho, \theta)$ variables,
\begin{equation}
\label{vvv}
{\cal{L}}=\rho\dot{\theta}-(\frac{1}{2}\rho (\pa_{i}\theta)^{2}+V(\rho))
\end{equation}
Next using (\ref{uuu}) and (\ref{vvv}), we obtain,
\begin{equation}
\label{ppp}
T^{i0}=\rho \pa^{i}\theta \dot{\theta}
\end{equation}
The $\dot{\theta}$ term is found by bracketing with the hamiltonian, exploiting the algebra (\ref{hii}),
\begin{equation}
\nonumber
\dot{\theta}=\{\theta, H\}=\{\theta, \int dy (\frac{1}{2}\rho (\pa_{i}\theta)^{2}+V(\rho))\}
\end{equation}
\begin{equation}
\label{abc}
=\frac{1}{2}(\pa_{i}\theta)^{2}+V_{\rho}(\rho)
\end{equation}
Thus,
\begin{equation}
\label{hih}
T^{i0}=\rho \pa^{i}\theta(\frac{1}{2}(\pa_{i}\theta)^{2}+V_{\rho}(\rho))=\rho v^{i}(\frac{1}{2}v^{2}+V_{\rho}(\rho))
\end{equation}
which is the familiar expression for the energy flux.

Likewise the expression for the stress tensor $T^{ij}$, following from (\ref{uuu}) and (\ref{vvv}), is given by\footnote{We use the mostly negative metric, $g^{ij}=-\delta^{ij}$},
\begin{equation}
\label{gig}
T^{ij}=\rho(\pa^{i}\theta)(\pa^{j}\theta)-{\cal{L}}g^{ij}=\rho v^{i}v^{j}+{\cal{L}}\delta^{ij}
\end{equation}
This may be simplified by writing ${\cal{L}}$ as,
\begin{equation}
\nonumber
{\cal{L}}=\rho\dot{\theta}-(\frac{1}{2}\rho v^{2}+V(\rho))
\end{equation}
Now we use (\ref{abc}) to get,
\begin{equation}
\label{uvw}
{\cal{L}}=\rho(\frac{1}{2}v^{2}+V_{\rho}(\rho))-(\frac{1}{2}\rho v^{2}+V(\rho))=\rho V_{\rho}-V
\end{equation}
so that we can express $T^{ij}$ in our familiar form,
\begin{equation}
\label{def}
T^{ij}=\rho v^{i}v^{j}+(\rho V_{\rho}-V)\delta^{ij}
\end{equation}
Thus (\ref{vvv}) may be regarded as the lagrangian in Euler variables for an irrotational fluid. 

We now show that the usual hydrodynamical equations follow from(\ref{vvv}) which is equivalent to \eqref{eee} with the identification \eqref{iit}. Variation with respect to $\theta$ gives,
\begin{equation}
\label{xoc}
\dot{\rho}-\pa_{i}(\rho(\pa_{i}\theta))=0
\end{equation}
If we now use(\ref{iit}), the above equation reproduces the continuity equation(\ref{pppp}).

Likewise, variation of $\rho$ in(\ref{vvv}) yields,
\begin{equation}
\label{xod}
\dot{\theta}-\frac{(\pa_{i}\theta)^{2}}{2}-V_{\rho}(\rho)=0
\end{equation}
Taking a spatial derivative and again exploiting(\ref{iit}) yields,
\begin{equation}
\label{xoe}
\dot{v_{i}}-v_{j}\pa_{i} v_{j}-\pa_{i}V_{\rho}=0
\end{equation}
For an irrotational fluid, $\pa_{i} v_{j}=\pa_{j} v_{i}$, so that(\ref{xoe}) just reproduces the Euler equation(\ref{rrrr}).

This may be extended to the general (non-vanishing vorticity) case, by an appropriate modification of the velocity parametrisation \eqref{iit}. It is not possible to do this by introducing just a single scalar like $\pa_{i}\beta$ or $\beta\pa_{i}\beta (=\frac{1}{2}\pa_{i}\beta^{2})$ since they can be absorbed in the original definition \eqref{iit}. Thus the simplest nontrivial possibility is to introduce a pair of scalars $(\alpha, \beta)$ and express the velocity as,
\begin{equation}
\label{pqr}
v^{i}=\partial^{i} \theta +\alpha \partial^{i}\beta
\end{equation}
and the hamiltonian now has the structure,
\begin{equation}
\label{chu}
H=\int dx (\frac{1}{2}\rho (\partial_{i} \theta +\alpha \partial_{i}\beta)^{2}+V(\rho))
\end{equation}
To see that this yields a meaningful fluid hamiltonian it is useful to construct the lagrangian with a suitable kinetic term. This is done in a way that reproduces the momentum density (\ref{cur}) with the velocity given by (\ref{pqr}), following the Noether prescription (\ref{uuu}). After  a little algebra we get the cherished form of the Lagrangian,
\begin{equation}
\label{che}
{\cal{L}}= \rho (\dot{\theta}+ \alpha \dot{\beta})-(\frac{\rho}{2} (\partial_{i} \theta +\alpha \partial_{i}\beta)^{2}+V(\rho))
\end{equation}
To verify our previous statements we compute the momentum density,
\begin{equation}
\label{bcd}
T^{0i}=\rho(\partial^{i} \theta +\alpha \partial^{i}\beta)
\end{equation}
which reproduces (\ref{pqr}) from (\ref{cur}).

Next, the energy flux $T^{i0}$ is computed from (\ref{uuu}) and(\ref{che}),
\begin{equation}
\label{lab}
T^{i0}=\rho (\dot{\theta}+ \alpha \dot{\beta})(\partial^{i} \theta +\alpha \partial^{i}\beta)
\end{equation}
The time derivatives may be eliminated by considering the equation of motion obtained by a variation of $\rho$ in (\ref{che}),
\begin{equation}
\label{lala}
\dot{\theta}+ \alpha \dot{\beta}=(\frac{1}{2} (\partial_{i} \theta +\alpha \partial_{i}\beta)^{2}+V_{\rho}(\rho))
\end{equation}
Substituting in (\ref{lab}) yields,
\begin{equation}
\label{hiji}
T^{i0}=\rho v^{i}[\frac{v^{2}}{2}+V_{\rho}(\rho)]
\end{equation}
where we have used (\ref{pqr}). Thus the desired form of the energy flux is reproduced.

Finally the stress tensor$T^{ij}$ is considered. Again exploiting Noether's definition (\ref{uuu}) and using (\ref{che}) we find,
\begin{equation}
\label{ijk}
T^{ij}=\rho(\pa^{i}\theta+\alpha\pa^{i}\beta)(\pa^{j}\theta+\alpha\pa^{j}\beta)+{\cal{L}}\delta^{ij}\\=\rho v^{i}v^{j}+{\cal{L}}\delta^{ij}
\end{equation}
on recalling the definition (\ref{pqr}). The Lagrangian density (\ref{che}) simplifies, using (\ref{lala}), to the form,
\begin{equation}
\label{nl}
{\cal{L}}=\rho V_{\rho}(\rho)-V(\rho)
\end{equation}
which, together with (\ref{ijk}), reproduces the expected structure of the stress tensor $T^{ij}$. Incidentally the above equation is expected on general grounds since it expresses the equivalence of the Lagrangian density with the fluid pressure. 

Now the hydrodynamical equations are derived from (\ref{che}). Variation with respect to $\theta$ yields, 
\begin{equation}
\label{abcd}
\dot{\rho}-\pa_{i}[\rho(\pa_{i}\theta+\alpha\pa_{i}\beta)]=0
\end{equation}
Using the form of the velocity (\ref{pqr}) in the above equation immediately reproduces the continuity equation (\ref{pppp}).

The demonstration of the Euler equation (\ref{rrrr}) needs some more work. Variation of $\alpha$ yields,
\begin{equation}
\label{bcde}
\dot{\beta}=v_{i}\pa_{i}\beta
\end{equation}
while that of $\beta$ yields,
\begin{equation}
\label{cdef}
\dot{\alpha}=v_{i}\pa_{i}\alpha
\end{equation}
where, at an intermediate step, the continuity equation (\ref{pppp}) has been used. Now, time differentiating the relation(\ref{pqr}), we find,
\begin{equation}
\label{defg}
\dot{v_{i}}=v_{j}\pa_{i}v_{j}+(v_{j}\pa_{j}\alpha)\pa_{i}\beta-(v_{j}\pa_{j}\beta)\pa_{i}\alpha + \pa_{i}V_{\rho}
\end{equation}
where all time derivatives appearing on the r.h.s of (\ref{pqr}) have been eliminated by exploiting (\ref{lala}), (\ref{bcde}) and (\ref{cdef}). The final point is to calculate the vorticity from (\ref{pqr}),
\begin{equation}
\label{efgh}
\omega_{ij}=\pa_{i}v_{j}-\pa_{j}v_{i}=
\pa_{i}\alpha\pa_{j}\beta-
\pa_{j}\alpha\pa_{i}\beta
\end{equation}
and use it to replace $\pa_{i}v_{j}$ in favour of $\pa_{j}v_{i}$ in (\ref{defg}). Immediately this reproduces the Euler equation (\ref{rrrr}).

We have thus shown that \eqref{che} may be regarded as the lagrangian density for vortical fluids. The decomposition \eqref{pqr} is the usual Clebsch parametrisation of a vector in terms of three scalars. However we have not put this by hand as had been done earlier. Rather, it was systematically derived.

Let us now consider the hamiltonian formalism. For this it is necessary to know the algebra of the basic variables. This may be obtained from (\ref{che}), by noting that $(\rho, \theta)$ and $(\rho\alpha, \beta)$ are the independent  canonical pairs. Then it is possible to to show that, apart from (\ref{hii}), the only other nonvanishing algebra among the basic variables is given by,
\begin{equation}
\label{hihi}
\{ \beta (x) ,\alpha (x') \}=\frac{1}{\rho}\delta (x-x'), ~~~\{ \alpha (x) ,\theta (x') \}=\frac{\alpha}{\rho}\delta (x-x')
\end{equation}
with all other brackets  being zero.

Now the complete algebra (\ref{two}) for a fluid with vorticity may be verified using (\ref{pqr}) and the brackets (\ref{hii}, \ref{hihi}). Consequently the fluid equations are also reproduced. This completes the hamiltonian analysis the Eulerian fluid model. 

It may be observed that the non canonical algebra (\ref{two}) was directly obtained in the hamiltonian formalism either as a 
postulate or by using generalised coordinates\cite{gre, sarlo, mor}, or by using the map connecting Lagrange to Euler variables. Here they simply follow from the modified symplectic structure.

\section{Nonisentropic fluids}

So far isentropic fluids were considered where entropy has no role. However, for a complete characterisation of a fluid, apart from its density and velocity, one has to include entropy. This section is devoted to a study of nonisentropic fluids along the lines developed in the earlier section.

The fundamental fluid equations are now the continuity equation, the Euler equation and the entropy convection. While the continuity equation (\ref{pppp}) remains unchanged, the Euler equation (\ref{rrrr}) becomes 
\begin{equation}
\label{enfu}
\dot{v_{i}}=v_{j}\pa_{j}v_{i}+\pa_{i}V_{\rho}(\rho,S)-\frac{\pa_{i}S}{\rho}V_{S}(\rho,S)
\end{equation}
where the potential is now a function of both the density and entropy $V(\rho, S)$. Here $S$ is the entropy per unit mass or the specific entropy. The above equation may be expressed in a more conventional form \cite{gre} by introducing the variable $U(\rho, S)$ as $V=\rho U$. Then (\ref{enfu}) reduces to, 
\begin{equation}
\label{zaa}
\dot{v_{i}}=v_{j}\pa_{j}v_{i}+ \frac{1}{\rho}\pa_{i}(\rho^{2}U_{\rho})
\end{equation}
which is the general form of the hydrodynamic force balance equation or the Euler equation.

Finally, the entropy convection equation, expressing the fact that heat flow is assumed to vanish, is given by 
\begin{equation}
\label{zbb}
\dot{S}=v_{i}\pa_{i}S
\end{equation}
Also the form of the Hamiltonian remains unchanged except that the potential is now a function of both $\rho$ and $S$,
\begin{equation}
\label{zcc}
H=\int(\frac{\rho v^{2}}{2}+V(\rho, S))
\end{equation}

We now discuss an action principle following our  earlier prescription. The idea is to modify the lagrangian so  that the momentum density is given by \eqref{cur}. This  obviously implies that the parametrisation for the velocity \eqref{pqr} has to be generalised. Following our earlier logic 
discussed below \eqref{xoe}, this may be done in two possible ways,
\begin{equation}
\label{zdd}
v^{i}=\pa^{i}\theta +\alpha\pa^{i}\beta +S\pa^{i}\gamma
\end{equation}
or, alternatively,
\begin{equation}
\label{zee}
v^{i}=\pa^{i}\theta +\alpha\pa^{i}\beta +\gamma\pa^{i}S
\end{equation}
It is interesting to note that both these forms lead to a consistent formulation. While the first representation (\ref{zdd}) was found earlier \cite{sarl, sel} using contact transformations,the second one (\ref{zee}) is new.
Let us first consider the analysis with (\ref{zdd}). To construct the Lagrangian corresponding to the Hamiltonian (\ref{zcc}), the kinetic term has to be defined. As mentioned it is done in a way that reproduces the momentum density (\ref{cur}) with the velocity given by (\ref{zdd}), using Noether's definition (\ref{uuu}). We find, 
\begin{equation}
\label{zff}
{\cal{L}}=\rho (\dot{\theta}+\alpha\dot{\beta}+S \dot{\gamma})-(\frac{\rho}{2}(\pa^{i}\theta +\alpha\pa^{i}\beta +S\pa^{i}\gamma)^{2}+ V(\rho, S))
\end{equation}
For a check we compute $T^{0i}$ from (\ref{uuu}),
\begin{equation}
\label{zgg}
T^{0i}=\rho v^{i}=\rho (\pa^{i}\theta +\alpha\pa^{i}\beta +S\pa^{i}\gamma)
\end{equation}
which immediately yields (\ref{zdd}) from (\ref{cur}). 

It is now possible to reproduce the expected structure of the flux
\begin{equation}
\label{zoo}
T^{i0}=\rho v^{i}[\frac{v^{2}}{2}+V_{\rho}(\rho , S)]
\end{equation}
 and the stress tensor (\ref{ijk}) from Noether's definition (\ref{uuu}). This proves the validity of the Lagrangian (\ref{zff}).

 The equations of motion for all the variables may be obtained from (\ref{zff}) by appropriate variations. Variation with respect to $\theta$ just yields the continuity equation(\ref{pppp}) once the velocity is identified as (\ref{zdd}). This is exactly as happened in the isentropic theory(\ref{che}). The $\gamma$ variation gives,
 \begin{equation}
 \label{ijkl}
 \pa_{t}(\rho S)-\pa_{i}(\rho S v_{i})=0
 \end{equation}
 where we have used (\ref{zdd}). A simple use of the continuity equation (\ref{pppp}) now reproduces the entropy convection equation (\ref{zbb}).
 
 The derivation of the Euler equation (\ref{enfu}) follows along the earlier isentropic case. First, the equations obtained on varying $\rho$, $S$, $\alpha$, $\beta$ are found to be, respectively,
 \begin{equation}
 \label{jklm}
 \begin{aligned}
 \dot{\theta}+\alpha\dot{\beta}+S\dot{\gamma}&=\frac{v^{2}}{2}+V_{\rho}\\
 \dot{\gamma}&=v_{j}\pa_{j}\gamma +\frac{V_{S}}{\rho}\\
 \dot{\beta}&=v_{i}\pa_{i}\beta\\
 \dot{\alpha}&=v_{i}\pa_{i}\alpha
 \end{aligned}
 \end{equation}
 
 The last two equations are, expectedly, identical to (\ref{bcde}) and (\ref{cdef}). Now time differentiating(\ref{zdd}) and eliminating the time derivatives appearing on its r.h.s by using (\ref{jklm}), we find,
 \begin{equation}
 \label{klmn}
 \dot{v_{i}}= v_{j}(-\pa_{i}\alpha \pa_{j}\beta -(\pa_{i}S)\pa_{j}\gamma +(\pa_{j}\alpha)\pa_{i}\beta +\pa_{j}S\pa_{i}\gamma + \pa_{i}v_{j}) + \pa_{i} V_{\rho}-\frac{V_{S}}{\rho}\pa_{i}S
 \end{equation}
 
 Computing vorticity from(\ref{zdd}),
 \begin{equation}
 \label{lmno}
 \pa_{i}v_{j}-\pa_{j}v_{i}=\pa_{i}\alpha\pa_{j}\beta +\pa_{i}S \pa_{j}\gamma -\pa_{j}\alpha\pa_{i}\beta -\pa_{j}S \pa_{i}\gamma
 \end{equation}
and substituting in (\ref{klmn}) reproduces the cherished equation(\ref{enfu}).
 
 For a hamiltonian analysis the noncanonical brackets have to be derived. These are obtained by  first noting that an inspection of \eqref{zgg} immediately identifies  the independent canonical pairs as, $(\rho , \theta)$ $(\rho\alpha ,\beta)$ and $(\rho S, \gamma)$. Here we compute the brackets involving $S$ since the others have already been found. It is easy to see that the only nonvanishing brackets are given by,
\begin{align}
\label{zkk}
& \{S(x),\gamma (x')\} =-\frac{1}{\rho}\delta(x-x'),~~~\{S(x),\theta (x')\} =\frac{S}{\rho}\delta(x-x'), 
\end{align}
 The bracket of $S$ with $v^{i}$ is now evaluated using(\ref{zkk}), and \eqref{zdd},
\begin{align}
\label{zll}
\{S(x), v^{i}(x')\} &=\{S(x), (\pa^{i}\theta +\alpha\pa^{i}\beta +S \pa^{i}\gamma)(x')\} \notag \\ & =\pa^{i}_{x'}(\frac{S}{\rho}\delta(x-x'))+S(x')\pa^{i}_{x'}(-\frac{1}{\rho}\delta(x-x')) \notag \\ & =\frac{\pa^{i}S}{\rho}\delta(x-x')
\end{align}

Using this algebra the equation \eqref{zbb}  is reproduced by bracketing S with  the hamiltonian \eqref{zcc}. Likewise the Euler equation \eqref{enfu} may also be reproduced.
Let us next consider the parametrisation (\ref{zee}). In this case the Lagrangian will be given by,
\begin{equation}
\label{znn}
{\cal{L}}=\rho (\dot{\theta}+\alpha\dot{\beta}+\gamma \dot{S})-(\frac{\rho}{2}(\pa^{i}\theta +\alpha\pa^{i}\beta +\gamma \pa^{i}S)^{2}+ V(\rho, S))
\end{equation}
The relevant equations (\ref{enfu}) and (\ref{zbb}) for the velocity and entropy, respectively, are reproduced from the above Lagrangian. For instance, varying with $\gamma$ immediately yields (\ref{zbb}).

The hamiltonian formulation may be performed as before by noting the difference in the canonical pairs from the earlier case. These pairs are now $(\rho , \theta)$ $(\rho\alpha ,\beta)$ and $(\rho\gamma ,S)$. With this change the algebra of $S$ is modified as, 
\begin{equation}
\label{zqq}
\{S(x),\gamma (x')\} =\frac{1}{\rho}\delta(x-x'),~~ \{S(x),\theta (x')\}=0
\end{equation}
while the other brackets remain unaltered. Also, the important bracket of $S$ with $v^{i}$ remains the same,
\begin{align}
\label{zrr}
\{S(x), v^{i}(x')\} &=\{S(x), (\pa^{i}\theta +\alpha\pa^{i}\beta +\gamma\pa^{i}S)(x')\} =\frac{\pa^{i}S}{\rho}\delta(x-x')
\end{align}
so that the equation of motion (\ref{zbb}) for the entropy is reproduced.

The fact that both parametrisations (\ref{zdd}, \ref{zee}) yield a consistent formulation is linked to the result that the algebra among the basic fluid variables $(v^{i}, \rho,S)$ is preserved. Thus the fundamental fluid equations like the continuity equation , the Euler equation and the entropy convection equation are all intact. Also, the conservation of energy momentum complex, which is a consequence of these equations, holds in either parametrisation.

Finally, let us discuss the freedom in choosing the potentials $\theta$, $\alpha$ etc that appear in the definition of the velocity (\ref{zdd}) or (\ref{zee}). Two sets of velocity potentials will be considered physically equivalent if they give the same velocity and also preserve the bracket structure. Let's start with two different sets of scalar variables $(\theta ,\alpha ,\beta , \gamma)$ and $(\theta ^{'} ,\alpha ^{'} ,\beta ^{'} , \gamma ^{'})$.
We have the condition
\begin{equation}
\label{fred}
v^{i}=\pa^{i}\theta +\alpha\pa^{i}\beta +\gamma\pa^{i}S=\pa^{i}\theta ' +\alpha '\pa^{i}\beta ' +\gamma '\pa^{i}S
\end{equation}
From this equality we can write,
\begin{equation}
\label{dine}
\pa^{i}(\theta -\theta')=\alpha\pa^{i}\beta - \alpha '\pa^{i}\beta ' -\pa^{i}S(\gamma -\gamma ')
\end{equation}
Now, two scalars $\theta$ and $\theta '$ differ by another scalar $F$. From \eqref{dine},
\begin{equation}
\label{oloo}
\pa^{i}F=\alpha\pa^{i}\beta - \alpha '\pa^{i}\beta ' -\pa^{i}S(\gamma -\gamma ')
\end{equation}
Now, different choice of $F$ will generate different transformation. We are interested only in the bracket preserving transformation.
For, in that case, the Hamiltonian as well as the equations of motion will be sustained. Instead of providing a general solution\footnote{For the relativistic theory this has been discussed in\cite{sch} where, however, the bracket structure is not highlighted.}, we discuss a specific example which is physically motivated.

Equivalent sets must have the same $S$ since entropy is a physical variable. Among the other variables, an inspection of (\ref{zkk}) reveals that the only vanishing bracket exists between $\theta$ and $\gamma$. This allows us to keep $\theta$ and $\gamma$ unchanged and modify the $\alpha$, $\beta$ sector. Defining the new potentials as,
\begin{equation}
\label{zvv}
\quad \theta ' =\theta, \quad \gamma '=\gamma, \quad \beta '=f(\beta), \quad \alpha '= \alpha(\frac{df}{d \beta})^{-1}
\end{equation}
where $f(\beta)$ is some function of $\beta$. It is then simple to verify,
\begin{equation}
\label{zyy}
v_{i}'=\pa_{i}\theta ' +\alpha '\pa_{i}\beta '+\gamma' \pa^{i}S'=v_{i}
\end{equation}
We can also find the generator of this transformation. A little bit of calculation shows, 
$$F= \alpha '(\beta '- f(\beta))$$
Also, the entire algebraic structure of \eqref{zkk} (along with \eqref{hihi})  is preserved. As an example,
\begin{equation}
\label{example}
\left\{ \alpha'(x), \theta'(x') \right\} = \left\{ \alpha \left( \frac{df}{d\beta} \right)^{-1} (x), \theta(x') \right\} =  \left(\frac{df}{d\beta}\right)^{-1} \frac{\alpha}{\rho} \, \delta(x - x')= \frac{\alpha'}{\rho} \, \delta(x-x)
\end{equation}
Preserving the velocity \eqref{zyy} implies preserving the hamiltonian. Since the algebra \eqref{zkk} is also invariant, it means that the fluid equations remain unchanged. Moreover the particular solution \eqref{zvv} shows that only one of the potentials (say $\beta$) may be completely arbitrary. This agrees with our intuitive notion of the fluid degrees of freedom. A fluid may be described by four functions at a point(one thermodynamic variable S and three independent components of the velocity). Since we used five potentials to define the fluid, only one of them should be completely arbitrary.

\section{Schwinger-type relations}
 We next consider the algebra among the components of the energy momentum tensor $T^{\mu\nu}$. Note that while $T^{ij}$ is symmetric since rotational invariance is preserved, $T^{0i}$ is not equal to $T^{i0}$ since space and time are on inequivalent footings in the non-relativistic theory. Strictly speaking, therefore, $T^{\mu\nu}$ should be referred as energy-momentum `complex'.

Recalling the algebra (\ref{two}) and the identification (\ref{cur}), it is possible to obtain, 
\begin{equation}
\label{scw}
\{T^{0i}(x), T^{0j}(x')\}=T^{0j}(x)\pa^{i}\delta (x-x')+T^{0i}(x')\pa^{j}\delta (x-x')
\end{equation}

This is the  typical form for one of the Schwinger conditions valid in the relativistic field theory. Indeed this structure is inbuilt in the very framework of the Eulerian fluids characterised by the algebra (\ref{qqq}) and the identification (\ref{cur}). Let us therefore look at the $T_{00}-T_{00}$ algebra that enters in the fundamental Schwinger condition. After some steps, using (\ref{two}), one obtains,
\begin{align}
\label{scc}
\{T_{00}(x), T_{00}(x')\} & =\{(\frac{1}{2}\rho v^{2}+V(\rho , S))(x),(\frac{1}{2}\rho v^{2}+V(\rho , S))(x')\} \notag \\& =(T_{i0}(x)+T_{i0}(x'))\pa_{i}\delta (x-x')
\end{align}
where $T_{i0}$ is defined in(\ref{zoo}). This is the famous Schwinger condition. A general proof is given in \cite{nger} for its validity in relativistic quantum field theory  where the Poisson or Dirac brackets are replaced by appropriate commutators. Interestingly such a relation was found by us \cite{ras} for relativistic
classical fluids. Here we  show its existence even for non relativistic classical fluids.

The existence of the Schwinger type condition (\ref{scc}) for nonrelativistic classical fluids is a new result. Since the proof was general, this condition is obviously valid for isentropic fluids also.

It is easy to see the connection of (\ref{scc}) with the conservation of the energy-momentum complex,
\begin{equation}
\label{jii}
 \pa_{\mu}T^{\mu \nu}=0
\end{equation}
Taking the integral  over space on both sides of (\ref{scc}), we find, 
\begin{equation}
\label{scd}
\{T_{00}(x),\int dx' T_{00}(x')\}=\int dx' (T_{i0}(x)+T_{i0}(x'))\pa_{i}\delta (x-x')
\end{equation}
which simplifies  to,
\begin{equation}
\label{sce}
\{T_{00}(x), H\}= \pa_{i}T_{i0}(x)
\end{equation}
on dropping  surface terms. Since  the l.h.s of the above equation is $\pa_{0}T_{00}$, we reproduce the time component of (\ref{jii}),
\begin{equation}
\label{fkj}
\pa_{\mu}T^{\mu 0}=0.
\end{equation}

The presence of Schwinger-like conditions gives a fresh insight into the interpretation of fluid dynamics as a field theory. Apart 
from revealing a very close correspondence with field theoretical results it can find various applications. One such instance was the derivation of 
(\ref{fkj}). Other possibilities include, for example, the demonstration of the Galilean algebra.

\section{Conclusion}

To conclude, we have given a new approach to construct the Lagrangian of a non relativistic Eulerian fluid. It is well known that, although the  hamiltonian  formulation follows along conventional lines, there are obstructions to the Lagrangian formalism. These are a consequence of the presence of Casimirs that prevent the inversion of the symplectic matrix so that the usual passage from the Hamiltonian to the lagrangian poses problems. It therefore becomes mandatory to take recourse to other avenues. Lin and Eckart\cite{kara} solved the problem by exploiting conservation laws and replacing the fluid velocity in favour of the Clebsh variables\cite{cl}. Such an approach is somewhat riddled with ambiguities. The use of conservation laws is ad-hoc, particularly since one of this laws has no physical basis\cite{jac}. Also, the  introduction of the Clebsh parametrisation is arbitrary and not logically motivated.

Our approach is based on equating the known expressions for the hamiltonian density, momentum density, energy flux and the stress tensor with those found by Noether's definition. The fluid velocity then gets naturally expressed in a Clebsh form for the irrotational case. The general(non-vanishing vorticity) example is subsequently treated by a logical extension. Algebraic consistency of the method has also been established. Including the entropic effects does not pose problems. For the nonisentropic fluid two forms (\ref{zdd}, \ref{zee}) of Clebsh decompositions are given that are useful for discussing the canonical formulation.While one of these decompositions was found previously\cite{sarl, sel} through canonical or contact transformations, the other is a new finding. The physical results, however, remain unaffected whether $S$ occurs in a differentiated form or otherwise. This shows the robustness of the scheme.

We have also discussed the arbitrariness in the choice of potentials appearing in the Clebsh decomposition of the velocity. While preserving the definition of velocity, the algebraic structure is also retained. Since invariance of the velocity naturally leads to the invariance of the Hamiltonian, the hydrodynamical equations are all preserved. From this analysis a degree of freedom count was done that agreed with physical considerations.

A word about the bracket structure is useful. These noncanonical brackets involving the velocity potentials and fluid density were earlier given by using involved canonical transformation \cite{sarlo} or by inspection \cite{gre} or by using the Lagrange to Euler map \cite{avi}. A systematic derivation using only Euler variables seems to be lacking. Our action principle fills this gap. The noncanonical brackets were easily read-off from the symplectic structure. Since this algebra plays an important role in reproducing the hydrodynamical equations and establishing compatibility between the Hamiltonian and Lagrangian formulations, its systematic derivation is definitely desirable.

Another point of the paper is to reveal the occurrence of a Schwinger-type condition that involves the algebra of the hamiltonian densities $T^{00}-T^{00}$. It may be recalled that, for a consistent formulation of relativistic quantum field theory, such a condition was proved by Schwinger \cite{nger} on general grounds. In the case of classical fluids, we had earlier reported the existence of such a condition for the relativistic case \cite{ras}. Quite surprisingly we find that such a relation also exists for non-relativistic fluids. There is a slight subtlety involved here. In the relativistic example since  the energy momentum tensor must be symmetric,$T^{0i}=T^{i0}$. This is of course not valid in the non-relativistic theory. Thus one has to be careful in writing the  appropriate form that occurs in the r.h.s of (\ref{scc}). A consistency check of (\ref{scc}) was done. It was shown that the integrated version of this equation was just the time component of the conservation law (\ref{jii}). The appearance of the Schwinger-type relation for a non-relativistic fluid theory is both surprising and interesting. Apart from showing the utility of the Clebsh parametrisation it strongly suggests the possibility of extending this analysis for the relativistic case. Some attempts in this direction were made by us in \cite{ras} but only for an isentropic fluid. More complicated theories involving interactions may also be considered within this formalism. An analysis for relativistic fluids in the light cone variables could be an interesting future prospect. This is so because light cone reduction is used to describe the dimensional reduction of relativistic fluids to their nonrelativistic counterparts.

  As a final remark we mention that the present approach can be immediately applied to develop a relativistic theory of
  fluids by appropriately generalising the Clebsh parametrisations (\ref{zdd}, \ref{zee}). Similar applications to magnetohydrodynamics, nonrelativistic or relativistic, may be envisaged.


\begin{thebibliography}{99}
\bibitem{hub} V.E.Hubeny, S.Minwalla, M.Rangamani The fluid/gravity correspondence
(arXiv:1107.5780) [hep-th]
\bibitem{lan} L. D. Landau, E. M. Lifshitz, Fluid Mechanics (Course of Theoretical Physics - volume
6), Elsevier, UK (1959).

\bibitem{hol}  D. Holm, B.A. Kupershmidt,  Phys.Lett. {\bf{A 101}}, 23 (1984). 
\bibitem{gre}  P.J.Morrison, J.M.Greene, Phys. Rev. Lett. {\bf{45}}, 790(1980)
\bibitem{jac} R. Jackiw, V. P. Nair, S. -Y. Pi, A. P. Polychronakos, J.Phys. {\bf{A37}}, R327-R432,  (2004), 
\bibitem{avi} E.C.D'Avignon, P.J.Morrison, M.Lingam "Derivation of the Hall and Extended Magnetohydrodynamics Brackets,"  	arXiv:1512.00942 [physics.plasm-ph]
\bibitem{baz} D. Bazeia, R. Jackiw Ann. Phys., NY {\bf{270}}~246(1988) 
\bibitem{cl}A. Clebsh, J. Reine Angew. Math. 56, 1 (1859).
\bibitem{kara} C. Eckart, Phys. Rev. 54, 920 (1938); C.C. Lin, International School of Physics E. Fermi (XXI); G. Careri, ed. (Academic Press,New York 1963).
\bibitem{ff} H. Fukagawa, Y. Fujitani, Prog.\ Theor.\ Phys.\ 124 (2010) 517; ibid. 127 (2012) 921
\bibitem{koda} J. H. G. Elsas, T. Koide, T. Kodama 'Noether Theorem of Relativistic-Electromagnetic Ideal Hydrodynamics' (arXiv: 1411.3238).
\bibitem{nger} J.Schwinger, Phys. Rev.{\bf{127}}, 324(1962)
\bibitem{sarlo} W.V.Saarloos, D.Bedeaux, P.Mazur, Physica {\bf{107A}} (1981) 109.
\bibitem{mor}C. Chandre (CPT), P. J. Morrison (IFS), E. Tassi (CPT) On the Hamiltonian formulation of incompressible ideal fluids and magnetohydrodynamics via Dirac's theory of constraints. (arXiv:1110. 6891)[physics.plasm-ph]
\bibitem{ras} R. Banerjee, S. Ghosh, A.K. Mitra EPJC 75:207 (2015) 
\bibitem{sarl} W.V.Saarloos, Physica {\bf{108A }} 557(1981)
\bibitem{sel} R.L.Seliger, G.B.Whitham, Proc. Roy. Soc {\bf{A305}}, 1(1968)
\bibitem{sch} B.F.Schutz, Phys. Rev {\bf{D2}}, 2762(1970)

 
\end{thebibliography}
\end{document}